\documentclass[twocolumn,showpacs,preprintnumbers,amsmath,amssymb]{revtex4}
\usepackage{tabularx,graphicx}\begin{document}
\newcommand{\beq}{\begin{equation}}
\newcommand{\eeq}{\end{equation}}
\newcommand{\beqn}{\begin{eqnarray}}
\newcommand{\eeqn}{\end{eqnarray}}
\newcommand{\bmath}{\begin{subequations}}
\newcommand{\emath}{\end{subequations}}
\title{Pair production and ionizing radiation from superconductors }
\author{J. E. Hirsch }
\address{Department of Physics, University of California, San Diego\\
La Jolla, CA 92093-0319}
 
\begin{abstract} 
We show that an alternative theory of superconductivity recently proposed
(theory of hole superconductivity) leads to the surprising consequence that
real electron-positron
pair production will occur for superconductors larger than a critical size. High frequency radiation with frequencies up to $0.511MeV/\hbar$  is 
predicted to be emitted from superconductors 
out of equilibrium. Attention to the possibility of harmful consequences is called for.   
\end{abstract}
\maketitle 

\section{Introduction}

The theory of hole superconductivity proposes that charge asymmetry
is at the root of the phenomenon of superconductivity\cite{charge}, and that this
charge asymmetry manifests itself at a macroscopic level in the fact that superconductors expel
negative charge from their interior towards the surface\cite{charge2}. No definitive experimental evidence for or
against this prediction exists so far, however the otherwise unexplained 'Tao effect'\cite{tao} has been argued
to constitute strong evidence in its favor\cite{tao2}. Because the theory of hole superconductivity is at odds with the
generally accepted BCS-London theory of conventional superconductivity\cite{schrieffer}, it is important to find unambiguous experimental evidence for or against it.

At a qualitative level the theory predicts that superconducting bodies look like 'giant atoms'\cite{giant}, with excess
negative charge near the surface and excess positive charge in the interior. The fact that the size of these
giant atoms is at the experimentalist's disposal then would appear to allow for a remarkably simple check of
the theory. Namely, it has been predicted that for superheavy atoms or molecules spontaneous electron-positron
pair production will occur\cite{atoms}, when the binding energy of a K-shell electron becomes equal to twice
its rest mass. Hence we are led to the surprising conclusion that 
 pair production should also occur  for superconductors of sufficiently large size.

Furthermore, the theory predicts that spin currents exist in the ground state of superconductors\cite{spinc}, with some electrons 
moving at speeds approaching the speed of light in macroscopic samples\cite{darwin}. 
  These electrons should give rise to high frequency radiation
when the superconducting state is destroyed and the spin current stops. 
In addition, annihilation of electron-positron pairs under suitable non-equilibrium 
conditions should give rise to 0.511 MeV $\gamma-$ray emission, that should be experimentally detectable.

We  call attention to the fact that if the theory is indeed correct, this ionizing radiation   could constitute a dangerous health hazard for humans
in the vicinity. Thus the determination of whether the theory is correct or not acquires an urgency that goes
 beyond the academic interest of deciding between competing theoretical viewpoints, since with increasing
use of superconductors in society this unexpected effect could have potentially harmful consequences.

\section{Pair production in superheavy atoms}

Shortly after the introduction of Dirac theory it was pointed out by Sauter\cite{sauter} that in a sufficiently strong
electric field electron-positron pair creation from the vacuum should take place. 
For an atom, an electron in the field of a point nucleus of charge $Z|e|$ has binding energy
\beq
E_b=-13.6 eV Z^2
\eeq
neglecting relativistic effects. It is generally believed that when the binding energy becomes degenerate with the
Dirac continuum of negative energy states, i.e. $E_b=-m_e c^2$, spontaneous pair production will occur\cite{atoms}
($m_e=$electron mass). From
Eq. (1), the condition for this to occur is $Z>274$. In fact, when relativity is taken into account the $1s$ state
for a point nucleus becomes unstable already for $Z>Z_c=137$ in Dirac theory. This can also be simply seen
from a Bohr atom model\cite{bohr}, using the relativistic equations
\bmath
\beq pv=\frac{Ze^2}{r}\eeq
\beq
p=\gamma m_e v
\eeq
\emath
$(\gamma=1/\sqrt{1-v^2/c^2})$ which together with angular momentum quantization $pr=\hbar$ lead to the condition
\beq
\frac{pc}{\sqrt{p^2c^2 + m_e^2 c^4}}=\frac{Ze^2}{\hbar c}
\eeq
so that for $Z=Z_c=\hbar c /e^2=137$, $p\rightarrow \infty$ and $r\rightarrow 0$. 
Solution of the Dirac equation for a nucleus of finite size shows that at a critical $Z_c=172$ spontaneous
autoionization will occur for an empty shell\cite{atoms}: an electron-positron pair pops out of the Dirac sea, the electron
will occupy the lowest orbit and the positron is emitted. Experimentally it has been attempted to reach
the supercritical region $Z>Z_c$ by colliding two heavy nuclei, however results to date have not been
conclusive\cite{atoms}. 

\section{Critical radius of superconductors}

\begin{figure}
\resizebox{6.5cm}{!}{\includegraphics[width=7cm]{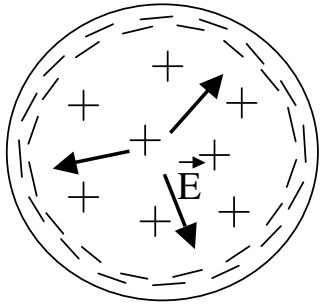}}
\caption{Schematic picture of charge distribution in a spherical superconductor of radius smaller than the critical radius.  Excess negative charge exists within a London penetration depth of the surface, and excess positive charge in the interior.}
\label{figure1}
\end{figure}

\begin{figure}
\resizebox{6.5cm}{!}{\includegraphics[width=7cm]{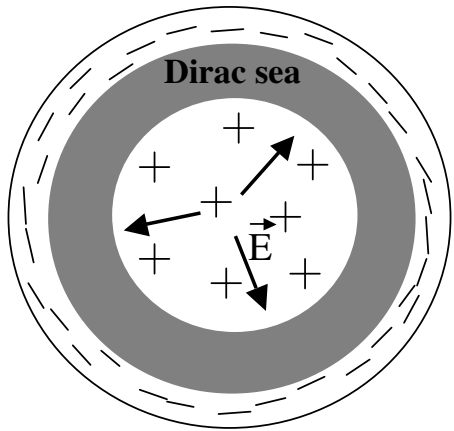}}
\caption{Schematic picture of charge distribution in a spherical superconductor of radius larger than the critical radius.  An intermediate region
between the inner positively charged region and outer negatively charged layer exists, where electron-positron pairs
pop out of the Dirac sea.}
\label{figure1}
\end{figure}

In a similar vein we argue that in superconductors spontaneous pair production will become possible for
superconductors larger than a critical size, within the theory of hole superconductivity. As discussed 
in \cite{charge2}, the theory predicts that electrons carrying a total negative charge
\bmath
\beq
q=4 \pi R^2 \lambda_L \rho_{-}
\eeq
get expelled from the interior of the superconductor towards the surface,
giving rise to a negative charge density $\rho_-$ within a London penetration depth $\lambda_L$ of the
surface (for simplicity we assume a sample of spherical shape of radius $R$), as shown schematically in Figure 1. The charge density
$\rho_-$ is given by\cite{charge2}
\beq
\rho_-=e n_s (\frac{10}{3} \frac{\epsilon}{m_e c^2})^{1/2}
\eeq
\emath
with $n_s$ the superfluid density, $e$ the $negative$ electron charge, and $\epsilon$ the condensation energy per electron. As the 
size of the sample becomes larger the amount of expelled charge increases, and we argue that when 
the potential energy of an electron at the edge of the positive charge distribution becomes larger in magnitude
than twice the electron rest energy
\beq
\frac{qe}{r}=2m_e c^2
\eeq
pair production will become energetically favorable,
as shown schematically in Fig. 2. Using Eqs. (4)   the condition Eq. (5) gives
a critical radius
\beq
R_c=\lambda_L (\frac{6m_e c^2}{5\epsilon})^{1/2}
\eeq

Alternatively, we may argue that the condition of dynamic equilibrium for an electron orbiting at radius r in the field of a positive charge $(-q)$
\beq
\frac{m_e v^2}{r}=\frac{qe}{r^2}
\eeq
yields that the speed $v$ approaches the speed of light $c$ for $qe/r=m_e c^2$, similar to Eq. (5). Of course the condition Eq. (7) ceases to be valid
for relativistic speeds and is replaced by Eq. (2) with $q=Ze$, which yields $v/c=0.91$ for $q$ given by Eq. (5).

As an example we consider Niobium. It was estimated in Ref. \cite{charge2} that the maximum electric field near the surface is 
$E_{max}\sim 0.77\times 10^6  V/cm$, which already indicates that for samples of size of order $cm$ the electron electrostatic energy becomes of the
order of the electron rest mass. The London penetration depth for $Nb$ is $\lambda_L=400 \AA$ and the thermodynamic critical field $H_c=1980G$,
from which  we find
condensation energy per unit volume $\tilde{\epsilon}=
1.56 \times 10^5 ergs/cm^3$,  $\epsilon=5.54 \mu eV$, and from Eq. (6)
\beq
R_c=3.33 \times 10^5 \lambda_L
\eeq
hence $R_c=1.33 cm$.  

\section{Dynamical equilibrium in applied magnetic field}

In the previous section we argued that for sufficiently large superconductors the electric field resulting from negative charge expulsion
will lead to pair production. Here we show that this expectation is consistent with the Meissner effect and the requirement of
dynamical equilibrium for superfluid electrons.

We assume the validity of London's equation
\beq
\vec{\nabla}\times\vec{J}=-\frac{c}{4\pi\lambda_L^2}\vec{B}
\eeq
for the screening charge current $\vec{J}$ in the presence of an applied magnetic field $\vec{B}$, with $\vec{J}$ given by
\beq
\vec{J}=\rho\vec{v}_\phi .
\eeq
Here, $\rho$ is the superfluid transport charge density and $\vec{v}_\phi$ its azimuthal velocity induced by the applied magnetic field. In 
conventional London theory it is assumed that $\rho=en_s$, with $n_s$ the total superfluid charge density, independent of the volume of
the sample, however we show here that this is untenable in the present context.

In the presence of the electric field $\vec{E}$
resulting from charge expulsion (Fig. 1), the expelled electrons near the surface will carry a spin current
even in the absence of an applied magnetic field\cite{spinc}, to satisfy dynamical equilibrium, with 
\beq
\frac{m_e v_0^2}{r}=|e|E
\eeq
for electrons at radius $r$ (neglecting relativistic corrections). Electrons of opposite spin orbit with opposite velocities of equal
magnitude $v_0$. When a magnetic field is applied, velocities of spin up and down electrons change according to
\bmath
\beq
\vec{v}_\uparrow=\vec{v}_0 + \vec{v}_\phi
\eeq
\beq
\vec{v}_\downarrow=-\vec{v}_0 + \vec{v}_\phi
\eeq
and in particular for equatorial orbits
\beq
v_\sigma=\sigma v_0+v_\phi
\eeq
\emath
so that the requirement of dynamical equilibrium for equatorial orbits is
\beq
\frac{m_e v_\sigma^2}{r}=|e|E+\frac{|e|}{c} v_\sigma B
\eeq
which   implies
\bmath
\beq
v_\phi=-\frac{e}{m_e c} \frac{Br}{2}
\eeq
The azimuthal velocity $v_\phi$ induced by the magnetic field Eq. (14a) is much larger than what would result if $\rho=en_s$ in Eq. (10), 
namely
\beq
v_\phi\sim-\frac{e}{m_e c}  B \lambda_L .
\eeq
\emath

From Eq. (9), since $B$ is non-zero only in a region of thickness $\lambda_L$ near the surface,
\beq
J\sim -\frac{c}{4\pi \lambda_L} B
\eeq
and from Eqs. (10), (14a) and (15)
\beq
\rho=\frac{m_e c^2}{2\pi \lambda_L e r}
\eeq
hence the transport charge density decreases inversely with the radius $r$. Using for the London penetration depth
\beq
\frac{1}{\lambda_L^2}=\frac{4\pi n_s e^2}{m_e c^2}
\eeq
Eq. (16) becomes
\beq
\rho=2en_s\frac{\lambda_L}{r}
\eeq
which shows that the transport charge density approaches the conventional value $en_s$ for small samples, but is much smaller than the
conventional value for samples much larger than the London penetration depth.

Now it is reasonable to assume that for  macroscopic samples the transport charge density cannot be smaller than the expelled charge
density $\rho_-$. From Eqs. (4a) and (16) we find, setting $\rho=\rho_-$ and $r=R$
\beq
\frac{qe}{R}=2m_e c^2
\eeq
i.e. the same condition as Eq. (5). We conclude that at the critical radius given by Eq. (6) the transport charge becomes equal to the
excess expelled negative charge Eq. (4).

\section{Interpretation of the Meissner effect}

Here we argue that the conclusion reached in the previous section that the transport charge in macroscopic samples is only the excess
charge $\rho_-$ rather than the full superfluid charge density $en_s$  also leads to an understanding of the Meissner effect.
Indeed, consider cooling a superconductor in the presence of an applied magnetic field $\vec{B}$. The charge that is expelled
from the interior towards the surface experiences a Lorentz force due to the magnetic field\beq
\frac{d\vec{v}}{dt}=\frac{e}{m_e c}\vec{v}\times\vec{B}
\eeq
and the azimuthal velocity builds up as electrons move radially outward due to this force. Using $\vec{v}=d\vec{r}/dt$ we find on
integrating Eq. (20)
\beq
\vec{v}_\phi (t=\infty)=\frac{e}{m_e c} [\vec{r}(t=\infty)-\vec{r}(t=0)]\times \vec{B}
\eeq
The charge expulsion process results in a total excess negative charge
\beq
q=4\pi R^2 \lambda_L \rho_-
\eeq
residing in the layer of thickness $\lambda_L$ at the surface. This negative charge moved outwards
upon cooling from above to below $T_c$,  from an initial spherical volume of
radius $(R-\lambda_L)$ to the spherical shell of inner radius $(R-\lambda_L)$ and outer radius $R$, as depicted in Figure 3. As shown
in Fig. 3, an electron initially near the center of the sphere (denoted as 1) moves from radius $r(t=0)\sim 0$ to
radius $r(t=\infty)\sim R-\lambda_L$, and in so doing acquires an azimuthal speed
\bmath
\beq
v_{1\phi}\sim -\frac{e}{m_e c} B R
\eeq
according to Eq. (21). This is in agreement with the speed obtained from the condition of dynamical equilibrium 
with a pre-existent $v_0$, Eq. (14a).
Instead, an electron initially at radius $(R-\lambda_L)$ (denoted by 2 in Fig. 3) ends up at the surface of
radius $R$, acquiring an azimuthal speed
\beq
v_{2\phi} \sim -\frac{e}{m_e c} B\lambda_L
\eeq 
\emath
according to Eq (21), which is of the same form as the 'classical' speed Eq. (14b).

\begin{figure}
\resizebox{6.5cm}{!}{\includegraphics[width=7cm]{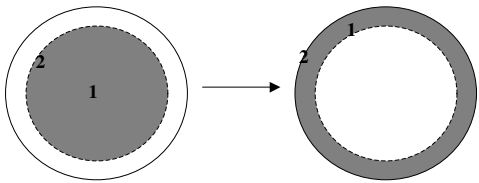}}
\caption{Schematic picture of the charge expulsion process. The shaded spherical negative charge distribution
of radius $R-\lambda_L$  on the left becomes
the outer shell of thickness $\lambda_L$ on the right. Electron at position 1, $r\sim 0$ on the left, moves to $r\sim R-\lambda_L$ on the
right; electron at position 2, $r\sim R-\lambda_L$ on the left, moves to $r\sim R$ on the right. }
\label{figure3}
\end{figure}

In summary, these arguments show that our conclusion in Sect. IV that electrons in the screening current have an excess
azimuthal speed $v_\phi$ which is very large as given by Eq. (14a) is consistent with the interpretation that these electrons are
expelled from the region near the center of the sphere, and acquire this azimuthal speed as their radial position changes
from $r\sim 0$ to $r\sim R-\lambda_L$ due to the Lorentz force. Instead, it is only electrons near the outer surface of the surface layer that move at the
'classical' speed Eq. (14b), since they only traversed a radial distance $\lambda_L$ in the process of charge expulsion.
In comparison, the conventional interpretation is that the entire superfluid density  contributes to  the Meissner screening
current with the classical speed Eq. (14b), however, that interpretation is incompatible with the 
requirement\cite{lorentz} that the azimuthal speed arises
from the Lorentz deflection of the outgoing radial motion.

\section{Interpretation of the London field of rotating superconductors}

In superconductors rotating with angular velocity $\vec{\omega}$ a uniform magnetic field \beq
\vec{B}=-\frac{2m_e c}{e}\vec{\omega}
\eeq
exists in the interior (London field)\cite{londonfield1,londonfield2,londonfield3}. This is conventionally interpreted as arising from a 'lagging' of the
superfluid rotation within a London penetration depth of the surface: superfluid electrons rotate slower than the rigid
rotation speed $v_\phi=\omega R \sin\theta$ by an amount\cite{londonfield2}
\beq
\delta v_\phi\sim\omega \lambda_L \sin\theta
\eeq
and the azimuthal current density is of order
\beq
J_\phi \sim e n_s \delta v_\phi \sim e n_s \omega \lambda_L
\eeq
The total current $I$ due to the superfluid electrons in a surface shell of thickness $\lambda_L$ is of order
\beq
I\sim 4\pi R^2 \lambda_LJ_\phi
\eeq
and the magnetic field due to a ring current of radius $R$ is of order 
\beq
B\sim \frac{2I}{cR}
\eeq
Replacing Eq. (27) in (28) and using Eq. (17) for the London penetration depth, the magnetic field Eq. (24) results.

The problem with this argument is that it does not provide a rationale for why the superfluid electrons near the surface
would suddenly 'lag behind' when a rotating normal metal is cooled into the superconducting state. Instead, our alternative
description does.

In our scenario the charge that 'lags behind' is the transport charge density Eq. (16), which for macroscopic samples is the
same as the excess charge density $\rho_-$ residing within a London penetration depth of the surface. Using
\beq
J_\phi =\rho \delta v_\phi
\eeq
together with Eqs. (16), (27) and (28) and demanding that the resulting magnetic field $B$ be the London field Eq. (24) leads to
\beq
\delta v_\phi \sim \frac{\omega R}{2} sin \theta
\eeq
If the transport charge density was expelled from the interior of the superconductor at an average radius $R/2$, it will lag
the azimuthal velocity at the surface by precisely the amount Eq. (30). Hence our point of view provides a simple rationale for
how the London field develops when a rotating normal metal is cooled into the superconducting state: the electrons near
the surface that suddenly 'lag behind' are electrons originating deep in the interior of the superconductor where much
smaller rigid rotation speeds prevail.

\section{Samples larger than critical: the intermediate layer}

Consider now a sample of radius $R$ that is larger than the critical radius $R_c$ given by Eq. (6). Let $q_1$ be the charge expelled from
the inner region, that satisfies
\beq
\frac{q_1 e}{R_c}=2m_e c^2 ,
\eeq
 and let $(q_1+q_2)$ be the total negative charge in the outer layer of thickness $\lambda_L$ that is responsible
for the charge transport. According to the condition of dynamical equilibrium derived in the previous section Eq. (19) we have
\beq
\frac{(q_1+q_2)e}{R}=2m_e c^2
\eeq
hence from Eqs. (31) and (32) we conclude that the electric potential is constant in the intermediate region $R_c<r<R$, 
consequently that {\it no electric field exists in the intermediate region}. Clearly, screening of the electric field has occurred through
creation of {\it real electron-positron pairs}. The positrons reside at the outer surface of the intermediate layer, and the
newly created electrons move out and add to the surface layer negative excess charge $\rho_-$, now satisfying
\beq
q_1+q_2=4 \pi R^2 \lambda_L \rho_{-}   .
\eeq
Hence the positive charge $(-q_2)$ is due to
$real$ $positrons$ created from the Dirac vacuum, and its magnitude is
\beq
|q_2|=|q_1| (\frac{R}{R_c}-1)
\eeq
with $q_1$ given by Eq. (4) with $R=R_c$ as given by Eq. (6).

For the case of $Nb$, $\rho_-=0.017 C/cm^3$\cite{charge2}, yielding $|q_1|=1.5\times 10^{-6}C$. For example,  in a sample of
radius $R=2R_C=2.66 cm$, Eq. (34)  predicts that   $9.4\times 10^{12}$ real electron-positron pairs created from the Dirac
vacuum exist!

Finally, we note that in the intermediate region there is no net electric field, and that because of the electron-positron pair creation the system
cannot be described with a wavefunction with a fixed number of particles, but rather requires a description that allows for
superposition of charge-neutral states with different numbers of particles. This is precisely the physical situation described
by the conventional BCS wave function.

\section{Pair production from externally applied electric field}
We have seen in the previous section that creation of real electron-positron pairs is expected to occur for large superconducting samples
when the internal electric field exceeds a critical value, to prevent the internal field from becoming larger. Here we show that
application of an external electric field is another mechanism leading to pair production.

Indeed, the relation between charge density $\rho(\vec{r})$ and electrostatic potential $\phi(\vec{r})$ in our theory is\cite{charge2}
\beq
\rho(\vec{r})-\rho_0=-\frac{1}{4 \pi \lambda_L^2} [\phi(\vec{r})-\phi_0(\vec{r})]
\eeq
where $\rho_0$ and $\phi_0(\vec{r})$ are the interior positive charge density and the associated electrostatic potential respectively.
Under an externally applied field the induced charge density is
\bmath
\beq
\rho_{ind}(\vec{r})=-\frac{1}{4\pi \lambda_L^2} \delta \phi(\vec{r}) \eeq
where $\delta \phi(\vec{r})$ is the change in the total electrostatic potential due to the applied electric field. Eq. (36a) shows that externally 
applied electric fields are screened over a London penetration depth, just as magnetic fields. Using Eq. (17) for $\lambda_L$,
\beq
\rho_{ind}(\vec{r})=-n_s e \frac{e\delta \phi(\vec{r})}{m_e c^2}
\eeq
\emath
In contrast, the induced charge density in  a normal metal when an external electric field is applied is
\bmath
\beq
\rho_{ind}(\vec{r})=-\frac{1}{4\pi \lambda_{TF}^2} \delta \phi(\vec{r}) \eeq
with the Thomas-Fermi screening length given by (for free electrons)
\beq
\frac{1}{\lambda_{TF}^2}=\frac{6 \pi n e^2}{\epsilon_F}
\eeq
with $n$ the electron density and $\epsilon_F$ the Fermi energy, so that Eq. (37a) is
\beq
\rho_{ind}(\vec{r})=-\frac{3}{2} n e \frac{e\delta\phi(\vec{r})}{\epsilon_F}
\eeq
\emath

Eq. (37c) for the normal metal shows that the fractional change in the local charge density $(ne)$ induced by the external field
is the ratio of the energy gain per electron $e\delta\phi$ to the energy cost in putting an extra electron at the top of the
Fermi distribution, $\epsilon_F$. Analogously, for the superfluid Eq. (36b) shows that the fractional change in the local
superfluid charge density induced by the external field is the ratio of $e\delta \phi$ to the energy cost in creating charges from
the Dirac vacuum, $m_e c^2$. Because this cost is much greater than $\epsilon_F$, the London length is much larger
than the Thomas Fermi length. We conclude that for the superfluid, screening of externally applied electric fields occurs through
pair production from the Dirac sea, because the 'rigidity' associated with the coherence of the superfluid wavefunction over
the macroscopic sample prevents screening through local shifts of the superfluid density.

Consequently, Eq. (36) directly furnishes the number of positrons created under application of an external electric field. The change in
potential under an applied electric field $E_{app}$ that decays to zero over a distance $\lambda_L$ is 
 \beq
\delta\phi\sim E_{app}\lambda_L \eeq
giving rise to an induced charge density
\beq
\rho_{ind}\sim -\frac{1}{4\pi \lambda_L} E_{app} \eeq
generated by pair production over a layer of thickness $\lambda_L$. The total charge created for a sample of surface area
$A$ exposed to the electric field is
\beq
q_{ind}\sim \rho_{ind}\lambda_L A=\frac{E_{app}}{4\pi}A
\eeq
hence the number of positrons created is
\beq
N_p\sim \frac{E_{app}A}{4\pi |e|}
\eeq
For example, for $E_{app}=1kV/mm$ and $A=1cm^2$, $N_p\sim 5.5\times 10^9$.
 
 \section{Observable consequences of pair production}
 We have seen in Sect. VI that in a spherical superconductor of radius $R$ larger than the critical radius $R_c$ given by
 Eq. (6), the amount of negative charge in the surface layer of thickness $\lambda_L$ exceeds the negative charge expelled
 from the interior by an amount
 \beq
 q_2=q(\frac{R}{R_c}-1)
 \eeq
 with $q$ given by Eq. (4). This excess negative charge originates in pair creation in the intermediate layer and reflects the
 existence of real positrons in the intermediate layer. Of course the same phenomenon is expected in large
 samples of non-spherical shape. Similarly we have argued in Sect. VII that when an external electric field is applied
 on a surface area $A$, real electron-positron pairs are created in a surface layer of thickness $\lambda_L$ to screen
 the applied field.
 
 When electrons and positrons collide they annihilate and two $\gamma-$rays of energy $m_e c^2=0.511 MeV$ are emitted in
 opposite directions. Of course in a superconductor in equilibrium or under stationary conditions no such  $\gamma-$ray emission
 is expected. 
 We propose however that under suitable non-stationary conditions $0.511MeV$ $\gamma-$rays will be emitted from superconductors, and
 suggest that an experimental effort to detect this effect should be undertaken. Because it is an unexpected effect within the
 conventional theory of superconductivity, if this phenomenon is detected it will shed important new light onto the physics
 of superconductivity.
 
 We suggest the following experimental tests. For a large superconducting sample at low temperatures, positrons should exist in 
 dynamical equilibrium with electrons in the intermediate layer. Rapid destruction of superconductivity by heating, or 
 application of ultrafast ultrastrong magnetic field pulses that drive the material normal, should cause electron-positron annihilation and 
 emission of $0.511MeV$ $\gamma-$rays that can be detected with appropriate detectors. Similarly, in the presence of an
 externally applied electric field positrons will exist which will annihilate and give rise to $\gamma-$ray emission if the 
 electric field is suddenly switched off. We suggest that placing a superconducting sample between the plates of a capacitor and
 increasing the electric field until breakdown occurs will give rise to $\gamma-$ray emission when the capacitor discharges
 through the superconductor. Alternatively, $\gamma-$ray emission should also occur upon application of 
 a strong rapidly varying ac electric field.
 
 However, $\gamma-$radiation can be hazardous to humans. Exposure to 5 rads per year is usually regarded as the limit
 of safety, and $5\times 10^9$ photons of energy $0.511MeV$ per square cm of tissue is $1$ rad. For the example
 discussed in Sect. VI, a sample of $Nb$ of radius $R=2R_c=2.66 cm$ has $\sim 10^{13}$ electron-positron pairs; if each
 pair emits $2$ photons when the sample goes normal it results in $2\times 10^{13}$ photons and a radiation dose of
 $0.13$ rads to a person $50$ $cm$ away. In cycling the sample from below to above $T_c$ several times, very
 quickly the limit of safety for this individual is met! For larger samples the danger becomes rapidly larger as seen
 from Eqs. (4) and (27). Similarly, application of large electric fields to superconductors can result in significant numbers of
 electron-positron pairs created, as discussed in Sect. VII, and experiments with large time-varying electric
 fields   could also result
 in hazardous amounts of $\gamma-$ radiation.

 \section{Bremsstrahlung}
 
 In addition to 0.511 MeV radiation, we expect that high energy radiation with a broad spectral range will be emitted from
 superconductors that are rapidly driven normal through sudden changes in temperature or applied magnetic field. 
 
  \begin{figure}
\resizebox{6.5cm}{!}{\includegraphics[width=7cm]{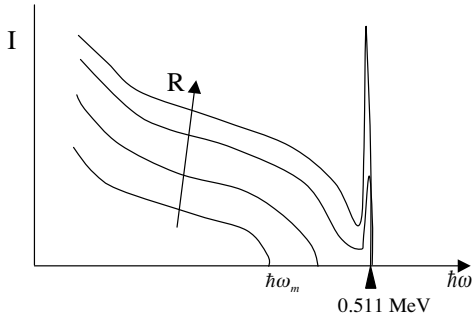}}
\caption{Schematic picture of the intensity of radiation versus frequency expected when a superconductor is heated from below to above $T_c$.
$R$ is the radius of the superconductor for a spherical sample. For $R$ larger than the critical radius $R_c$ (Eq. 6) the peak at 0.511MeV appears }
\label{figure4}
\end{figure}

 Indeed, the electrons expelled from the interior that give rise to the outer negative charge density $\rho_-$, carry a spin current
 in the superconducting state. At radius $r$ the kinetic energy of these electrons is, from Eq. (7):
 \beq
 \frac{1}{2}m_e v_0^2=\frac{q e}{2r}
 \eeq
 where $q$ is the net charge inside $r$. In particular, the fastest speeds occur for electrons at the edge of the
 positive charge distribution shown in Fig. 1. If the superconductor is suddenly driven normal these electrons carrying the macroscopic
 spin current will stop and emit bremsstrahlung, of maximum frequency $\omega_m$ determined by conversion of the entire
 kinetic energy of the electron  into a single photon:
 \beq
 \hbar \omega_m(r)=\frac{qe}{2r}
 \eeq
 Hence at the critical radius given by Eq. (31) we obtain from Eq. (44)
 \beq
 \hbar \omega_m(R_c)=0.511 MeV
 \eeq
 which is the same as the photon energy resulting from electron-positron destruction. For radius $R$ smaller than the critical radius we have simply
 \beq
  \hbar \omega_m=\frac{R}{R_c}m_e c^2
  \eeq
 Consequently we expect a broad spectrum of high energy 
 radiation, with the upper limit frequency $\omega_m$ increasing with sample size: samples of radius smaller than $2\times 10^{-4}R_c$ will emit in
 the UV  ($\hbar \omega_m<100eV$), samples of radius up to $R=0.2R_c$ will also emit X-rays  ($100eV<\hbar \omega_m<100keV)$
 and samples with $R>0.2R_c$ will in addition emit $\gamma-$rays ($\hbar \omega_m>100keV$), up to a maximum
 frequency $511 keV/\hbar$ when the radius reaches the
 critical radius  $R_c$.
 The radiation will
 originate predominantly from the region at distance $\lambda_L$ from the surface of the sample, where the fastest electrons
 in the spin current reside. When the system becomes normal, these 'undressed' electrons\cite{holeelec2} in Cooper pairs will suddenly unbind
 and experience scattering by the discrete ionic potential, and emit a bremsstrahlung spectrum 
 as given by the Bethe-Heitler formula\cite{bethe}.  The  spectral distribution detected will depend both on the bremsstrahlung processes and
 on the scattering processes that occur in the path of the photon towards the surface. When the sample radius becomes
 larger than $R_c$, a peak will grow at the maximum frequency $511 keV/\hbar$ reflecting the electron-positron
 annihilation processes. This is schematically depicted in Fig. 4.

The integrated intensity of the bremsstrahlung radiation should be proportional to the negative charge in the outer layer shown
in Fig. 1. Hence we expect the integrated intensity to grow proportionally to $R^2$ for samples of radius smaller than
$R_c$, according to Eq. (4). For $R>R_c$, the negative charge density $\rho_-$ starts to decrease according to Eq. (16), so 
the integrated intensity (excluding the peak at $0.511 MeV$) should grow proportionally to $R$. The peak at $0.511MeV$ should increase
proportionally to $(R/R_c-1)$ according to Eq. (34).

We conclude from these considerations that experiments with and practical uses of large superconducting samples, as well as
 processes involving application of large electric fields to superconductors, are potentially dangerous.  The level of
ionizing radiation generated in these situations should be ascertained before they can be safely carried out in an 
 environment where humans are in danger of exposure.

\section{Conclusions}
Superconductivity has been traditionally regarded as a low energy phenomenon, because low temperatures are involved and
because in the conventional theory phonons, that are low energy excitations in the solid, are thought to play the
dominant role. Only recently, evidence from optical experiments in high $T_c$ superconductors\cite{marel} has suggested that
 higher energy scales, in the mid-infrared and visible range (of order $eV$) play a role at least in those materials. We have also
  recently suggested 
changes in the plasmon dispersion relation\cite{electro}, which for conventional superconductors can be
above $10 eV$;  this prediction has not yet been put to experimental test.
Continuing this trend, in this paper we have suggested that energy scales relevant to superconductivity extend even much
higher, to the range of millions of $eV$.

How can $MeV$ energies possibly be relevant for a phenomenon where the local energies involved are of order $\mu eV$, i.e. a factor
$10^{12}$ smaller? Qualitatively, the key lies in the quantum coherence of the superconducting state over macroscopic
distances. In a sample of volume $1 cm^3$ there are of order $10^{23}$ atoms in the bulk, and of order $10^{18}$ atoms in the
surface layer of thickness $\lambda_L$. Hence a fraction $10^{-7}$ of the surface layer atoms could each display a
phenomenon at an energy scale of $10^6 eV$ if they are able to harness an energy of $10^{-6} eV$ from each of the
atoms in the bulk. Therein lies the remarkable nature of the macroscopic quantum coherence that is the hallmark of 
superconductivity.

The importance of relativity in the theory of hole superconductivity was already foreshadowed early on  in the
lattice formulation of the theory\cite{hole}, which describes pairing and superconductivity as driven by an off-diagonal 
Coulomb interaction term in the Hamiltonian, $\Delta t$. This term gives rise to a $reduction$ $of$ $the$ $mass$
of the carriers when they form a Cooper pair\cite{london}, evidence for which has recently been seen experimentally\cite{marel}.
In relativity a $bound$ $state$ of two particles necessarily has a smaller mass than the sum of its
constituent's masses due to the energy-mass relation $E=mc^2$.

The work discussed here also sheds new light on the meaning of the BCS wave function. The fact that the BCS wave function
describes the superconducting state as a superposition of states with different number of particles has until now been regarded
merely as a convenient calculational device, without physical content. Indeed, there is no $physical$ reason in the conventional
theory for why a wave function describing pairing of electrons could not be described with a fixed number of pairs. Furthermore
there is something profoundly unphysical about the BCS wave function in the conventional context: each Cooper pair carries
a mass of $1.022 MeV/c^2$ and a charge of $2e$, so that the BCS wavefunction superposes states with widely different
electric charges and energies. Why would the description of a low-energy phenomenon require the mixing of such very different states?
Instead, in the present context the superposition of states with different number of electrons and positrons (but the same
total electric charge) arises as a $consequence$ $of$ $the$ $physics$ and indicates that a BCS-like wavefunction that 
superposes different occupation number sectors is in fact $required$ to describe the physical reality.

In physics, the first ``hole theory'' proposed was that of Dirac\cite{dirac}, to deal with the negative energy states that
he encountered in formulating the relativistic quantum theory of the electron. In this paper we have shown that
the theory of hole superconductivity leads unavoidably to the inclusion of Dirac's holes (positrons) in the description of 
the superconducting state.

The theory discussed here predicts that ionizing radiation with a continuum of frequencies all the way up to
$0.511 MeV/\hbar$ will be emitted from large superconductors in 
non-equilibrium processes. No other theory of superconductivity predicts this effect, hence detection of such radiation
will support the theory of hole superconductivity, or call for alternative explanations.

To conclude we emphasize again that in the process of advancing science, incomplete or erroneous understanding can be dangerous. 
Before the discovery of capacitors it was thought that the bigger an object the more electricity it could store; then, 
Musschenbrock described his experiment with a small Leyden jar with the words "suddenly I received in my right hand a shock of
such violence that my whole body was shaken as by a lightning stroke". William Roentgen died of bone cancer and Marie Curie 
of leukemia, presumably triggered by exposure to X-rays and radioactivity respectively in the course of performing their experiments.
If the widely accepted BCS-London theory is correct for conventional superconductors, and d-wave superconductivity for
high $T_c$ cuprates, no danger from harmful ionizing radiation from superconductors
should be expected; nevertheless, no matter how small the perceived chance to the contrary, I suggest that  it behooves scientists to rule out 
  the scenario proposed in this paper.

\acknowledgements 
I am grateful to  Joaquin Fernandez-Rossier and Frank Marsiglio for
stimulating comments.


\begin{references}
\bibitem{charge} J.E. Hirsch, Phys. Lett. A {\bf 134}, 451 (1989); cond-mat/0407642 (2004) and references therein.
\bibitem{charge2} J.E. Hirsch, Phys.Rev.B {\bf 68},184502 (2003).
\bibitem{tao} R. Tao, Physics World, August 2005, p.3 (http://physicsweb.org/articles/world/18/8/5/1); R. Tao , X. Xu , Y.C.  Lan , Y. Shiroyanagi , Physica C {\bf 377}, 357 (2002).
\bibitem{tao2} J.E. Hirsch, Phys. Rev. Lett.  {\bf 94}, 187001 (2005).
\bibitem{spinc}  J.E. Hirsch,  Phys.Rev. B{\bf 71}, 184521  (2005).
\bibitem{darwin} J.E. Hirsch,  Phys.Lett. A {\bf 345}, 453 (2005).
\bibitem{schrieffer} J.R. Schrieffer, ``Theory of Superconductivity'', 
Addison-Wesley Publishing Company, Redwood City, 1964.
\bibitem{giant} J.E. Hirsch, Phys.Lett. A {\bf 309}, 457 (2003).
\bibitem{atoms} B. Muller, Ann. Rev. Nucl. Sci. {\bf 26}, 351 (1976) and references therein.
\bibitem{sauter} F. Sauter, Z. Phys. {\bf 69}, 742 (1931).
\bibitem{bohr} S.S. Gershtein and Y.B. Zeldovich, Sov. Phys. JETP {\bf 30}, 358 (1970).
\bibitem{lorentz}  J.E. Hirsch, Phys.Lett. A {\bf 315}, 474 (2003).
\bibitem{londonfield1} R. Becker, F. Sauter and C. Heller, Z. Physik {\bf 85}, 772 (1933).
\bibitem{londonfield2} F. London, 'Superfluids', Dover, New York, 1961.
\bibitem{londonfield3} A.F. Hildebrand , Phys.Rev.Lett. {\bf 8}, 190 (1964).
\bibitem{holeelec2}  J.E. Hirsch, Phys.Rev. B{\bf 71},  104522  (2005).
\bibitem{bethe}  H. Bethe and W. Heitler, Proc. Roy. Soc. {\bf A146}, 83  (1934).
\bibitem{marel} H. J. A. Molegraaf, C. Presura, D. van der Marel, P. H. Kes, and M. Li
Science {\bf 295}, 2239 (2002);  A.F. Santander-Syro, R.P.S.M. Lobo, N. Bontemps, Z. Konstantinovic, 
Z.Z. Li and H. Raffy, Europhys.Lett.62, 568 (2003).
\bibitem{electro}  J.E. Hirsch, Phys.Rev. B{\bf 69}, 214515  (2004).
\bibitem{hole}  J.E. Hirsch and F. Marsiglio, Phys. Rev. B {\bf 39}, 11515 (1989); Physica C {\bf 162-164}, 591 (1989).
\bibitem{london} J.E. Hirsch and F. Marsiglio, Phys. Rev. B {\bf 45}, 4807 (1992);  J.E. Hirsch,  Physica C {\bf 199}, 305 (1992).
\bibitem{dirac} P.A.M. Dirac, Proc. Roy. Soc. London {\bf A126}, 360 (1929-30).

\end{references}
\end{document}